\documentclass[conference]{IEEEtran}
\IEEEoverridecommandlockouts
\usepackage{cite}
\usepackage{multirow} 
\usepackage{amsmath,amssymb,amsfonts}
\usepackage{lipsum,multicol}
\usepackage{algorithmic}
\usepackage{graphicx}
\usepackage{algorithmic}
\usepackage[ruled,vlined]{algorithm2e}
\SetKwInput{KwInput}{Input}  
\SetKwInput{KwOutput}{Output}              
\SetKw{Break}{break}
\SetKwRepeat{Do}{do}{while}
\usepackage{textcomp}
\usepackage{float}
\usepackage{xcolor}
\usepackage{stfloats}
\usepackage{empheq}

\def\BibTeX{{\rm B\kern-.05em{\sc i\kern-.025em b}\kern-.08em
    T\kern-.1667em\lower.7ex\hbox{E}\kern-.125emX}}

\begin{document}
\title{Max–Min Fairness in Stacked Intelligent Metasurface-Aided Rate Splitting Networks} 
\vspace{-10mm}
\author{ Abdullah Quran, Shimaa Naser, Maryam Tariq, Omar Alhussein, and Sami Muhaidat\\
\normalsize	 Department of Computer and Information Engineering, Khalifa University\\
 Email: \{100062547, shimaa.naser, 100036622, omar.alhussein, sami.muhaidat\}@ku.ac.ae 

\vspace{-4mm}
}
\maketitle

\begin{abstract}

This paper investigates a downlink multiuser multiple-input single-output system that integrates rate-splitting multiple access (RSMA) with a stacked intelligent metasurface (SIM) to enable wave-domain beamforming. Unlike conventional digital beamforming, the proposed system leverages the programmable phase shifts of the SIM to perform beamforming entirely in the wave domain. This work introduces a fairness-centric SIM-RSMA design that ensures equitable resource allocation, which is critical for applications such as industrial IoT and multi-user extended reality. We formulate a max-min rate optimization problem that jointly optimizes the base station’s transmit power and the SIM’s phase shifts. Due to the non-convex and coupled nature of this problem, we develop an alternating optimization framework where the power allocation is optimized via successive convex approximation and the SIM beamforming is optimized using the Riemannian conjugate gradient method. Simulation results demonstrate that the proposed SIM-RSMA framework achieves superior minimum user rates and enhanced fairness compared to SIM-assisted non-orthogonal multiple access schemes.

\end{abstract}

\vspace{-4mm}

\section{Introduction}

With the increasing demand for higher data rates and more energy-efficient communication infrastructure, traditional beamforming techniques rely heavily on digital signal processing and radio frequency (RF) chains are complex and costly for multiuser environments. Reconfigurable intelligent surfaces (RISs) have emerged as a transformative technology, enabling programmable control over electromagnetic (EM) wave propagation via passive meta-atom arrays \cite{ris_lit1}. RISs enhance spectral efficiency and energy efficiency by applying tunable phase shifts on incident waves \cite{ris_lit6}. However, conventional RISs are typically limited to single-layer reflection, acting as passive relays with limited spatial processing capabilities, which constrains their performance in complex wireless environments. To address these challenges, stacked intelligent metasurfaces (SIMs), inspired by multi-layer diffractive optics \cite{sim_lit2}, have been recently proposed to enable complex beamforming operations in the wave domain using multiple stacked metasurface layers, thereby reducing the reliance on digital beamforming hardware \cite{sim_lit3, sim_lit4}. 

Likewise, rate-splitting multiple access (RSMA~\cite{rsma0}) has emerged as a powerful and flexible strategy for multiuser interference management. RSMA splits each message into a common part decoded by all users and private parts decoded individual users. Hence, RSMA enables partial interference decoding, yielding superior spectral efficiency compared to conventional schemes such as space-division multiple access (SDMA~\cite{sdma}) and non-orthogonal multiple access (NOMA~\cite{rsma0}) 


The integration of RSMA with SIM  presents a promising direction for next-generation wireless networks.
SIM enables high-precision beamforming, suppressing inter-user interference via focused beam alignment. This fine-grained spatial control complements the interference management strategy of RSMA, which mitigates residual interference through message splitting.  Furthermore, SIM’s analog architecture reduces energy consumption and hardware complexity by eliminating the need for digital beamforming.
Combining SIM and RSMA form a robust and synergistic framework for improving performance and reducing cost and power in multiuser interference-limited and dense wireless environments.

Despite these advantages, the integration of SIM and RSMA remains unexplored. Recent studies have either explored the integration of RSMA with RISs~\cite{IRS_RSMA1}, which is constrained by the limited beamforming resolution of single-layer RISs, or focused on SIM for single-user systems and conventional multiple access~\cite{sim_lit3, sim_multi_user2}. A notable exception is the work presented in~\cite{sim_rsma} which considers SIM-RSMA in the wave domain but focuses on sum-rate maximization. This approach has critical limitations for multiuser systems where ensuring a uniform quality of service (QoS) is paramount, such as in industrial IoT sensor networks, autonomous vehicles, and multiuser extended reality applications~\cite{fairness_applications}. In these scenarios, the sum-rate objective fails to ensure rate allocation, often collapsing to a trivial solution that allocates most resources to users with the strongest channels, leading to resource deprivation for the remaining users.

In this work, we propose a max-min fairness framework for wave-domain SIM-RSMA systems that ensures balanced performance across all users. The proposed design effectively combines SIM’s fine-grained beamforming with RSMA’s interference management to enhance robustness and guarantee uniform rate QoS in multiuser environments.
However, achieving such fairness is significantly more challenging than sum-rate oriented designs as the optimization problem becomes highly non-convex and tightly coupled, involving joint power allocation and multi-layer phase-shift beamforming under strict unit-modulus constraints. Furthermore, the beamforming design becomes more sophisticated, requiring the SIM to prioritize users with weak channels and manage interference inclusively. This leads to interference-rich conditions that fully exploit RSMA’s potential: the common stream mitigates interference for all users, while the SIM’s beamforming focuses private streams to reduce mutual interference. The main contributions of this work are summarized as follows:

\begin{itemize}

    \item We propose the first max-min fairness design for the SIM-RSMA system, where wave-domain beamforming and RSMA-based message splitting are jointly optimized to maximize the minimum user rate. The formulated problem is highly non-convex due to coupled power and phase-shift variables and unit-modulus constraints.


    \item We develop an efficient alternating optimization (AO) framework to address this challenge. The power allocation and common rate are optimized via successive convex approximation (SCA), while the SIM phase shifts are updated using Riemannian conjugate gradient (RCG) on the complex circle manifold.

   \item We evaluate the proposed SIM-RSMA scheme against SIM-SDMA and SIM-NOMA baselines. Results show that SIM-RSMA outperforms SIM-NOMA while achieving comparable performance to SIM-SDMA in terms of minimum user rate and fairness. Moreover, the use of multilayer SIM is shown to achieve better performance compared to conventional single-layer SIM structures.
\end{itemize}

\vspace{-3mm}

\section{System Model and Problem Formulation}

\begin{figure}[h]
\centering 
\includegraphics[width=0.35\textwidth,keepaspectratio]{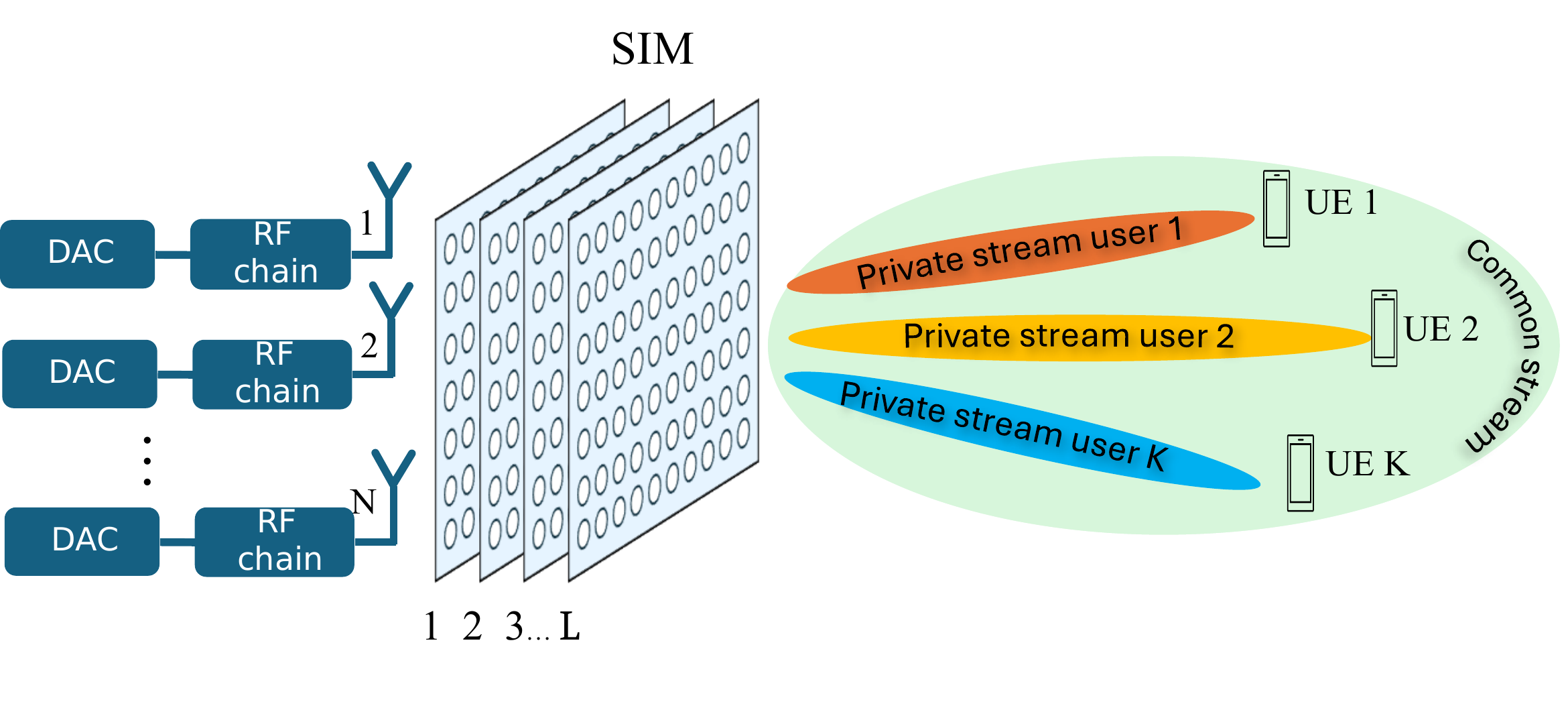}
\vspace{-0.5cm}
\caption{\footnotesize Illustration of the proposed SIM-assisted RSMA MISO system.}
\label{concept_figure}
\vspace{-0.2cm}
\end{figure}

As illustrated in Fig.~\ref{concept_figure}, we consider a downlink multiuser MISO system where a base station (BS) equipped with \( N \) antennas communicates with \( K \) single-antenna user equipments (UEs) via RSMA. In this work, we propose a SIM-based architecture to perform beamforming directly in the wave domain.  The SIM consists of \( L \) programmable metasurface layers with \( M \) meta-atoms per layer to perform beamforming directly in the EM domain. The transmitted signal consists of a common stream decodable by all UEs and private streams intended for individual UEs. Let $\mathcal{L} \!\!=\!\! \{1, 2, \ldots, L\}$, \!\!$\mathcal{M} \!\!=\!\! \{1, 2, \ldots, M\}$, and $\mathcal{K} \!\!=\!\! \{1, 2, \ldots, K\}$ represent the sets of metasurfaces, meta-atoms per layer, and UEs, respectively. The phase shift matrix corresponding to the \( \ell \)-th metasurface layer is denoted as \( \mathbf{\Theta}_\ell \!\!= \!\!\text{diag}\left(\theta_{\ell,1}, \theta_{\ell,2}, \ldots, \theta_{\ell,M}\right)
\in \mathbb{C}^{M \times M} \), $\forall$\( \ell \!\!\in\!\! \mathcal{L} \), where \( \theta_{\ell,m} \!\!= \!\!e^{j\phi_m^\ell} \) represents the EM response of the \( m \)-th meta-atom on the \( \ell \)-th metasurface layer, and \( \phi_m^\ell \in [0, 2\pi) \) denotes the corresponding phase shift. 
\vspace{-4mm}

The Rayleigh–Sommerfeld diffraction model \cite{sim_lit2} is employed to capture the near-field wave interactions between the metasurface layers, which is essential for modeling the wave-based beamforming capabilities of the multi-layer SIM. Specifically, the wave propagation from the \( (\ell - 1) \)-th to the \( \ell \)-th metasurface layer is modeled by a transmission matrix \( \mathbf{Q}_\ell \in \mathbb{C}^{M \times M} \), defined for all \( \ell \in \mathcal{L} \), \( \ell \neq 1 \). The \( (m, m') \)-th entry, \( q_\ell^{m, m'} \), of \( \mathbf{Q}_\ell \) represents the transmission coefficient from the \( m' \)-th meta-atom on the \( (\ell - 1) \)-th layer to the \( m \)-th meta-atom on the \( \ell \)-th layer, and is given by 
\begin{equation}
q_\ell^{m, m'} = \frac{A_t \cos \psi_\ell^{m, m'}}{d_\ell^{m, m'}} \left( \frac{1}{2\pi d_\ell^{m, m'}} - j\frac{1}{\lambda} \right) e^{j 2\pi d_\ell^{m, m'} / \lambda},
\label{eq:1}
\end{equation}
\noindent where \( \lambda \) is the wavelength, \( A_t \) is the area of each meta-atom, \( \psi_\ell^{m, m'} \) is the angle between the propagation direction and the surface normal, and \( d_\ell^{m, m'} \) is the corresponding propagation distance. The BS-to-SIM transformation matrix is denoted by \( \mathbf{Q}_1 = [\mathbf{q}_1, \mathbf{q}_2, \ldots, \mathbf{q}_N] \in \mathbb{C}^{M \times N} \), where each \( \mathbf{q}_n \in \mathbb{C}^{M \times 1} \) models the response from the \( n \)-th transmit antenna to the first metasurface layer. Following the convention in~\cite{sim_isac, sim_multi_user, sim_multi_user2, sim_sum_rate}, we assume each antenna is associated with one user. To implement 1-layer RSMA, the BS is equipped with \(N = K + 1\) antennas: \(K\) antennas transmit the private streams and the remaining antenna transmit the common stream. This configuration ensures independent control over both the transmit power and the achievable signal-to-interference-plus-noise ratio (SINR) for all stream, while granting the SIM sufficient spatial flexibility to manage multi-user interference through wave-based beamforming. The complete wave-domain response of the SIM is expressed as
\begin{equation}
\mathbf{F} = \mathbf{\Theta}_L \mathbf{Q}_L \mathbf{\Theta}_{L-1} \mathbf{Q}_{L-1} \cdots \mathbf{\Theta}_2 \mathbf{Q}_2 \mathbf{\Theta}_1 \in \mathbb{C}^{M \times M}.
\end{equation}

To model the end-to-end downlink channel from the SIM to each user, we assume a quasi-static flat-fading scenario. Let \( \mathbf{h}_k \in \mathbb{C}^{M \times 1} \) denote the effective channel vector from the SIM to the \( k \)-th user. The channel \( \mathbf{h}_k\) is assumed to follow a correlated Rayleigh fading model given by \( \mathbf{h}_k \sim \mathcal{CN}(\mathbf{0}, \zeta_k \mathbf{T}) \), where $\zeta_k$  represents the distance-dependent path loss and \( \mathbf{T} \in \mathbb{C}^{M \times M} \) is the spatial correlation matrix characterizing coupling between the meta-atoms. Under an isotropic multipath environment with a uniform power azimuth spectrum, the correlation between meta-atoms is represented by the Fourier transform of the uniform angular distribution. Thus, the correlation between the $m$-th and $m'$-th meta-atoms is modeled as $[\mathbf{T}]_{m,m'} = \text{sinc}\left(2 d_{m,m'}/\lambda\right)$, where $d_{m,m'}$ is the spacing between meta-atoms \cite{correlation}. Moreover, we assume perfect channel state information (CSI) of the SIM-User channels $\mathbf{h}_k$ is available at the BS.

Using the 1-layer RSMA strategy \cite{rsma0}, the BS transmits a signal that can be expressed as
$\mathbf{x} = \sqrt{p_c} s_c + \sum_{k \in \mathcal{K}} \sqrt{p_k} s_k, $
where \( s_c \sim \mathcal{CN}(0, 1) \) is the common message decoded by all users, and \( s_k \sim \mathcal{CN}(0, 1) \) is the private message intended for user \( k \). Accordingly, the received signal at user \( k \) is given by
\begin{equation} 
y_k = \mathbf{h}_k^H \mathbf{F} \left( \mathbf{q}_c \sqrt{p_c} s_c + \sum_{j \in \mathcal{K} } \mathbf{q}_j \sqrt{p_j} s_j \right) + n_k,
\end{equation} \noindent where \( n_k \sim \mathcal{CN}(0, \sigma_k^2) \) denotes the additive noise at user \( k \). Unlike conventional beamforming-based systems, the BS does not perform spatial precoding. Instead, the transmitted signal undergoes wave transformation within the SIM, where phase shifts across multiple layers control the signal propagation.

At the receiving side, each user applies successive interference cancellation (SIC) to decode its messages. The SINRs for decoding the common and private messages at user \( k \) are given by  
\begin{equation} 
\gamma_{\delta,k} = \frac{p_{\delta} |\mathbf{h}_k^H \mathbf{F} \mathbf{q}_\delta|^2}{\sum_{j \in \mathcal{K}_\delta(k)} p_j |\mathbf{h}_k^H \mathbf{F} \mathbf{q}_j|^2 + \sigma_k^2}, \quad \delta \in \{\mathrm{c}, \mathrm{p}\} \tag{4}
\end{equation} \noindent where \( \delta \in \{\mathrm{c}, \mathrm{p}\} \), and the index sets are  \( \mathcal{K}_\mathrm{c}(k) = \mathcal{K} \) and \( \mathcal{K}_\mathrm{p}(k) = \mathcal{K} \setminus \{k\} \). Accordingly, the achievable common and private rates for user \( k \) are given by \( R_{c,k} = \log_2(1+\gamma_{c,k}) \) and \( R_{p,k} = \log_2(1+\gamma_{p,k}) \), respectively. Since all users must successfully decode the common message, its rate is constrained by the weakest user, i.e., \( R_c \leq \min_k R_{c,k} \). Consequently, the individual achievable rate is given by $R_k = r_{c,k} + R_{p,k}$ such that $\sum_{k\in \mathcal K} r_{c,k}  \leq  R_{c}$, where \( \mathbf{r}_c = [r_{c,1}, r_{c,2}, \ldots, r_{c,K}]\) denotes the vector of auxiliary variables representing the common rate portions assigned to the users.

Aiming to maximize the minimum rate across all users, we formulate the joint optimization of the transmit power allocation and the phase shift configuration of the SIM as
\vspace{-1mm}
\begin{align}
(\text{P1}): & \max_{\mathbf{\Theta}_l, \mathbf{p}, \mathbf{r}_{c} } \quad  \min_{k \in \mathcal{K}} \{r_{c,k} + R_{p,k}\} \tag{5a}\\
& \quad \text{s.t.} \quad |\theta_{\ell,m}| = 1, \quad\quad\quad\quad \forall \ell \in \mathcal{L}, \forall m \in \mathcal{M}, \tag{5b}\\
& \quad \hphantom{\text{s.t.}} \quad p_c + \sum_{k \in \mathcal{K}} p_k \leq P_{\max},  \tag{5c}\\
&\quad \hphantom{\text{s.t.}} \quad \sum_{k\in \mathcal K} r_{c,k}  \leq  R_{c,k}, \quad r_{c,k} \geq  0,\quad \forall k\in \mathcal{K} \tag{5d}
\label{p1}
\end{align}
where \( \mathbf{p} = [p_c, p_1, \ldots, p_K]  \) is the vector of power allocation variables. Constraint (5b) imposes a unit-modulus condition on each phase shift element of the SIM.  constraint (5c) imposes a constraint on the power transmitted by the BS, ensuring it does not exceed $P_{\max}$, which represents the maximum power budget at the BS. Finally, constraint (5d) guarantees that the sum of the allocated common rates across all users remains within the decodable common rate at each user.

\vspace{-1mm}

\section{Joint Power Allocation and Wave-Based
Beamforming: Solution to Problem (P1)}

Problem (P1) is a non-convex and challenging optimization problem due to the coupled design of the transmit power variables and the SIM phase shifts in (5a), as well as the non-convex unit-modulus constraints in (5b). Therefore, obtaining the global optimal solution to (P1) is intractable. To address this, we propose an efficient algorithm based on alternating optimization to obtain a near-optimal solution for (P1). In particular, the AO framework alternately optimizes the power allocation and SIM phase shift variables. The corresponding power allocation subproblem and the SIM phase shift subproblem are introduced in subsections \ref{sub:PA} and \ref{sub:SIM}, respectively.

\vspace{-3mm}
\subsection{Power Allocation}
\label{sub:PA}
Given fixed SIM phase shift matrices \( \{ \boldsymbol{\Theta}_\ell \} \), we first optimize the power allocation vector \( \mathbf{p} \) and the common rate allocation vector \( \mathbf{r}_c \). The power allocation subproblem is formulated as
\vspace{-1mm}
\begin{align}
(\text{P2}): & \max_{\mathbf{p}, \mathbf{r}_{c}, t} \quad t \tag{6a} \\
& \quad \text{s.t.} \quad r_{c,k} + R_{p,k} \geq t, \quad \forall k \in \mathcal{K}, \tag{6b} \\
& \quad \hphantom{\text{s.t.}} \quad \text{(5c)}, \text{(5d)}. \tag{6c} \label{p2}
\end{align}
\noindent where $t$ is an auxiliary variable introduced to represent the minimum rate across all users and to enable a smooth reformulation of the original max–min objective. However, problem (P2) remains non-convex due to the non-convexity of constraint (5d) and constraint (6b), as both the private rates \( R_{p,k} \) and the common rates \( R_{c,k} \) are non-concave functions of the power allocation vector \( \mathbf{p} \).

We next express the achievable rates as the difference of two logarithmic terms to facilitate convexification. Specifically, the common and private rates are written as 
\begin{equation}
 R_{c,k} = \log_2(p_c |\mathbf{h}_k^H \mathbf{F} \mathbf{q}_c|^2 + \sum_{j \in \mathcal{K}} p_j |\mathbf{h}_k^H \mathbf{F}  \mathbf{q}_j|^2 + \sigma_k^2) - \eta_{c,k}, \tag{7}
\end{equation}
\begin{equation}
 R_{p,k} = \log_2( \sum_{j \in \mathcal{K}} p_j |\mathbf{h}_k^H \mathbf{F} \mathbf{q}_j|^2 + \sigma_k^2) - \eta_{p,k}, \tag{8}
\end{equation}
\noindent where $\eta_{c,k}$ and $\eta_{p,k}$ are expressed as follows 
\begin{equation}
\eta_{\delta,k} = \log_2\left( \sum_{j \in \mathcal{K}_\delta(k)} p_j |\mathbf{h}_k^H \mathbf{F} \mathbf{q}_j|^2 + \sigma_k^2 \right)\!, \quad \delta \in \{\mathrm{c}, \mathrm{p}\}.\tag{9}
\end{equation}

Despite this reformulation, \( R_{c,k} \) and \( R_{p,k} \) remain non-concave due to the non-concavity of the term \( \eta_{\delta,k} \). To address this, we approximate equation (9) using first-order Taylor expansions around the solution obtained at iteration \( t-1 \) as
\begin{equation}
\tilde{\eta}_{\delta,k}^{(t)} \!= \!\eta_{\delta,k}^{(t-1)} + \frac{ \sum_{j \in \mathcal{K}_\delta(k)} |\mathbf{h}_k^H \mathbf{F} \mathbf{q}_j|^2 (p_j - p_j^{(t-1)})}{\ln 2 \left( \sum_{i \in \mathcal{K}_\delta(k)} p_i^{(t-1)} |\mathbf{h}_k^H \mathbf{F} \mathbf{q}_i|^2 + \sigma_k^2 \right)}. \tag{10}
\end{equation}

\noindent Finally, \( R_{c,k} \) and \( R_{p,k} \) are now concave functions of the power allocation vector \( \mathbf{p} \). Consequently, constraints (5d) and (6b) are convex, and the approximate problem at each iteration is convex and can be efficiently solved using convex optimization solvers such as CVX \cite{cvx}. Hence, we employ the SCA method, which iteratively refines the solution by solving a sequence of convex subproblems. At iteration $t$, SCA approximates the non-concave components \( \eta_{c,k} \) and \( \eta_{p,k} \) with their first-order convex surrogates $\tilde{\eta}_{c,k}^{(t)}$ and $\tilde{\eta}_{p,k}^{(t)}$, evaluated using the solution from the previous iteration following equation (10). \vspace{-2mm}
\subsection{SIM Beamforming Optimization}
\label{sub:SIM}
Given a fixed power allocation $\mathbf{p}$, the SIM beamforming subproblem aims to maximize the minimum rate of all users by optimizing the phase shifts applied across all metasurface layers. The SIM beamforming subproblem is expressed as 
\begin{align}
(\text {P3}): & \max_{\mathbf{\Theta}_l, \mathbf{r} } \quad  \min_{k \in \mathcal{K}} \{r_{c,k} + R_{p,k}\} \tag{11a}\\
& \quad \text{s.t.} \quad \text{(5b)}, \text{(5d)}. \tag{11b}
\label{p3}
\end{align}

To facilitate the solution to (P3), we collect the diagonal elements of the phase shift matrices \( \{ \boldsymbol{\Theta}_\ell \}_{\ell=1}^L \) into a single stacked vector \( \boldsymbol{\theta} \in \mathbb{C}^{LM \times 1} \) defined as $\boldsymbol{\theta} = [\theta_{1,1}, \dots, \theta_{1,M}, \theta_{2,1}, \dots, \theta_{L,M}]^T$, where each element \( \theta_{\ell,m} \) satisfies the unit-modulus constraint (5b). 

The non-convex nature of constraint (5b) makes it particularly difficult to handle (P3) using traditional optimization techniques, necessitating more advanced solution methods. To handle the non-convexity of constraint (5b), we reinterpret the feasible set defined by (5b) as a Riemannian manifold\cite{book} $\mathcal{S}$ as follows
\begin{equation}
\mathcal{S} = \{ \boldsymbol{\theta} \in \mathbb{C}^{LM}: |\theta_i| = 1,  \quad \forall i \in \{1, \dots, LM\}, \tag{12}
\end{equation}
which corresponds to the Cartesian product of \( LM \) complex unit circles. This representation enables the use of Riemannian optimization techniques that respect the constraint (5b) and ensure feasibility at every iteration. The tangent space at a point \( \boldsymbol{\theta}\) to the manifold $\mathcal{S}$ is defined as $ T_{\boldsymbol{\theta}} \mathcal{S} = \{ \mathbf{z} \in \mathbb{C}^{LM} : \Re( z_i \theta_i^* ) = 0, \forall i \} $, where \( \Re(\cdot) \) denotes the real part.

Moreover, to address the non-smoothness of the objective function in (11a), we adopt a log-sum-exp surrogate to obtain a differentiable approximation as follows
\begin{equation}
\mathcal{F}(\boldsymbol{\theta}) = -\frac{1}{\beta} \log\left( \sum_{k \in \mathcal{K}} e^{-\beta (r_{c,k} + R_{p,k})} \right)\!,  \tag{13}
\end{equation}
where \( \beta > 0 \) is a parameter that controls the approximation tightness  and $\{r_{c,k}\}$ are fixed from the BS step.

Next, we account for the decodability constraint (5d). To do so, we introduce the violation measure
\begin{equation}
\Gamma(\boldsymbol{\theta}) = \sum_{k=1}^K r_{c,k} + \frac{1}{\beta}\log\!\left(\sum_{j=1}^K e^{-\beta R_{c,j}(\boldsymbol{\theta})}\right), \tag{14}
\end{equation}
which becomes positive whenever the constraint is not satisfied. This violation is then penalized through the differentiable function
\begin{equation}
\mathcal{F}_{\mathrm{pen}}(\boldsymbol{\theta}) 
= -\mu \cdot \frac{1}{\beta}\ln\!\big(1+e^{\beta \Gamma(\boldsymbol{\theta})}\big), \tag{15}
\end{equation}
where $\mu > 0$ determines the strength of enforcement. The penalty term drives the optimization away from infeasible points. While the penalty ensures feasibility, it may cause vanishing gradients once the constraint is satisfied. To maintain useful search directions in the strictly feasible region, we also introduce the common-stream margin
\begin{equation}
\Delta(\boldsymbol{\theta}) = -\frac{1}{\beta}\log\!\left(\sum_{j=1}^K e^{-\beta R_{c,j}(\boldsymbol{\theta})}\right) - \sum_{k=1}^K r_{c,k}, \tag{16}
\end{equation}
which quantifies the slack in the decodability constraint. Based on this margin, we define the reward term
\begin{equation}
\mathcal{F}_{\mathrm{rew}}(\boldsymbol{\theta}) 
= \eta \cdot \frac{1}{\beta}\ln\!\left(1+e^{\beta \Delta(\boldsymbol{\theta})}\right), \tag{17}
\end{equation}
with $\eta \ll \mu$ ensuring that feasibility remains the dominant objective. The reward encourages further improvement of the common-stream capacity once feasibility has been achieved.

By combining the fairness surrogate, penalty, and reward, the smooth SIM objective is given as
\begin{equation}
J(\boldsymbol{\theta}) 
= \mathcal{F}_{\mathrm{fair}}(\boldsymbol{\theta})
+ \mathcal{F}_{\mathrm{pen}}(\boldsymbol{\theta})
+ \mathcal{F}_{\mathrm{rew}}(\boldsymbol{\theta}). \tag{18}
\end{equation}
In this formulation, the penalty dominates whenever the decodability constraint is violated, while the reward contributes primarily in the feasible region. These two terms therefore act in a complementary fashion, ensuring both strict constraint satisfaction and continual improvement of the minimum achievable user rate.

This smooth reformulation facilitates the development of the RCG algorithm over the manifold \( \mathcal{S} \) to optimize the SIM phase shifts. In each iteration of the RCG algorithm, we compute the Euclidean gradient, $\nabla_{\boldsymbol{\theta}} \mathcal{J}$. Because updates based on the Euclidean gradient may not respect the manifold constraint $\mathcal{S}$, we project it onto the tangent space $T_{\boldsymbol{\theta}} \mathcal{S}$ to obtain the Riemannian gradient, expressed as follows

\begin{equation}
\mathbf{g}_{\boldsymbol{\theta}} = \nabla_{\boldsymbol{\theta}} \mathcal{J} - \Re\left( \nabla_{\boldsymbol{\theta}} \mathcal{J} \odot \boldsymbol{\theta}^* \right) \odot \boldsymbol{\theta}, \tag{19}
\end{equation}
where \( \odot \) denotes the element-wise product. The search direction at iteration \textit{t+1}, denoted by $\mathbf{u}^{(t+1)}$, which is conjugate to previous search directions, is then updated as follows:
\begin{equation}
\mathbf{u}^{(t+1)} = \mathbf{g}^{(t+1)} + \gamma^{(t+1)} \mathbf{u}^{(t)},\tag{20}
\end{equation}
where $\gamma^{(t+1)}$ is computed using the Polak–Ribière formula~\cite{book}. Subsequently, a retraction step is applied to project the new point back onto the manifold $\mathcal{S}$:
\begin{equation}
\theta_i^{(t+1)} = \frac{\theta_i^{(t)} + \alpha u_i^{(t)}}{|\theta_i^{(t)} + \alpha u_i^{(t)}|}, \quad \forall i \in \{1, \dots, LM\}.\tag{21}
\end{equation}
where \( \alpha \) is the step size determined via Armijo backtracking line search~\cite{book}. 
The RCG algorithm iteratively refines the phase shift vector \( \boldsymbol{\theta} \) until convergence, ensuring that the unit-modulus constraint (5b) is satisfied at every step.

The proposed algorithm alternates between SCA-based power allocation and RCG-based SIM beamforming. The power allocation subproblem involves solving a convex program with \( \mathcal{O}(K^3) \) complexity per SCA iteration, converging in \( t_1 \) steps. The SIM beamforming subproblem optimizes a length-\( LM \) phase vector shift using RCG, with complexity \( \mathcal{O}((LM)^{2}) \), converging in \( t_2 \) iterations \cite{book}. Thus, the total complexity over \( t \) AO iterations is \( \mathcal{O}(t(t_1 K^3 + t_2 (LM)^{2})) \).

\vspace{-2mm}

\section{Simulation Results}
\label{sec:results}

\begin{figure}[h]
\vspace{-6mm}
\centering 
\includegraphics[width=0.28\textwidth,keepaspectratio]{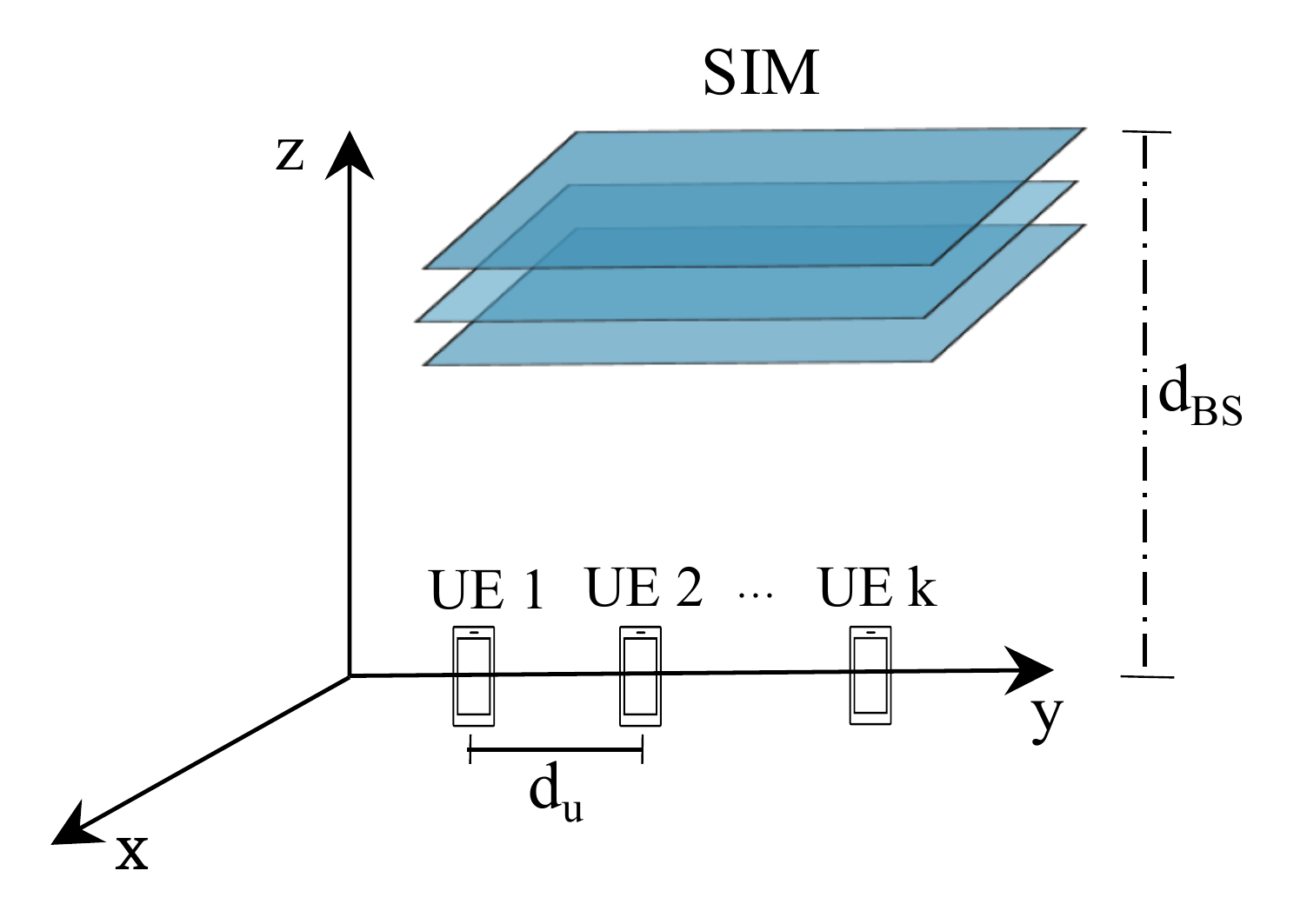}
\vspace{-0.5cm}
\caption{\footnotesize Simulation setup of the proposed SIM-assisted RSMA system.}
\label{fig:simulation_setup}
\vspace{-2mm}
\end{figure}

In this section, numerical simulations are provided to evaluate the performance of the proposed SIM-RSMA system. The BS, positioned at a height of $d_{BS}\!=\!5$ m, is equipped with N antennas arranged along the x-axis. The SIM’s layered structure lies parallel to the x-y plane, with all antennas or meta-atoms aligned along the z-axis, ensuring coordinated beam control across the aperture. Moreover, $K$ users are evenly spaced on the y-axis by $d_{u}$ of 5 m as illustrated in Fig. \ref{fig:simulation_setup}. The system operates at a carrier frequency of 28\,GHz. The SIM consists of $L$ metasurface layers, each with $M$ meta-atoms of area $A_t = \lambda^2/4$ arranged in a square grid of size $\sqrt{M} \times \sqrt{M}$. The spacing between adjacent meta-atoms within each layer is set to $\lambda/2$, and the total SIM thickness is $5\lambda$, with an inter-layer spacing of $5\lambda / L$. The wave propagation distance $d_{\ell}^{m,m'}$ between the $m'$-th meta-atom on layer $(\ell-1)$ and the $m$-th meta-atom on layer $\ell$ is therefore the Euclidean distance based on their relative spatial coordinates. The noise power at each user is $\sigma_k^2 \!=\! -100\,\text{dBm}$, and the distance-dependent path loss is modeled as $\zeta_k \!=\! \beta_0 (d_k/d_0)^{-\alpha}$, where $\beta_0 \!=\! -60\,\text{dB}$, $d_0 \!=\! 1\,\text{m}$, and $\alpha \!=\! 3.5$ \cite{sim_multi_user}. Unless otherwise mentioned, the parameters used in the simulation are as follows: the number of SIM elements per layer is $M\!=\!64$, the number of users $K\!=\!3$, and the power budget $P_{\max} = 20$dBm. 

We evaluate the proposed SIM-RSMA scheme against four benchmark schemes: (1) \textbf{SIM-SDMA} that applies private message transmission only, with optimized power allocation and SIM beamforming, (2) \textbf{SIM-NOMA} that uses power-domain multiplexing with SIC decoding \cite{rsma0}, assisted by SIM-based beamforming, (3) \textbf{RSMA} with fully digital beamforming, and (4) \textbf{SIM-RSMA-RND} that evaluates SIM-RSMA scheme with optimized power allocation and randomly phase shifts

\begin{figure*}[b]
\vspace{-5mm}
     \centering
     \begin{minipage}{0.28\textwidth}
         \centering
         \includegraphics[width=\textwidth]{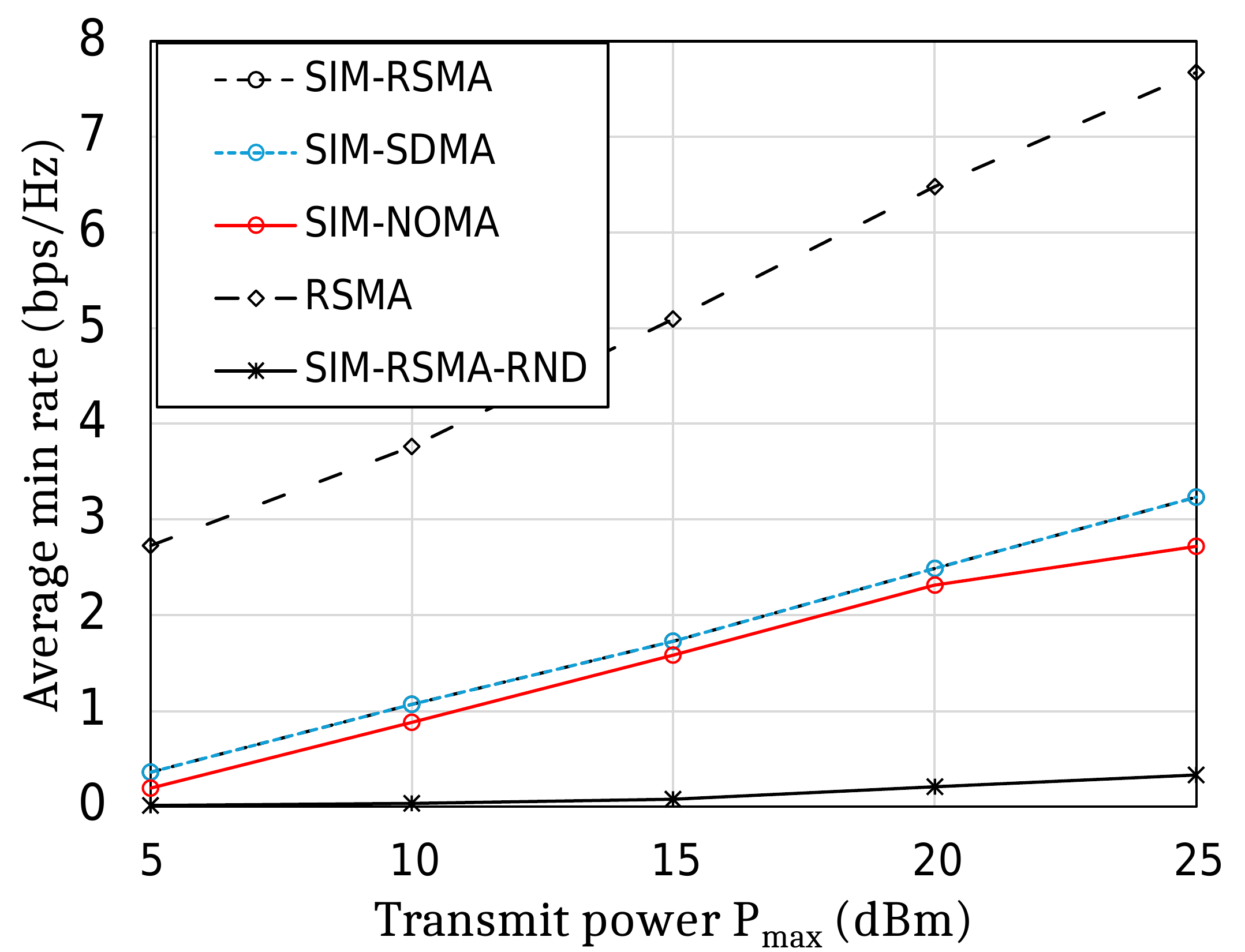}
        \vspace{-0.3in}
         \caption{The impact of the transmit power $P_{\max}$ on the minimum rate.}
         \label{fig:pmax}
     \end{minipage}
     \quad
     \begin{minipage}{0.28\textwidth}
         \centering
         \includegraphics[width=\textwidth]{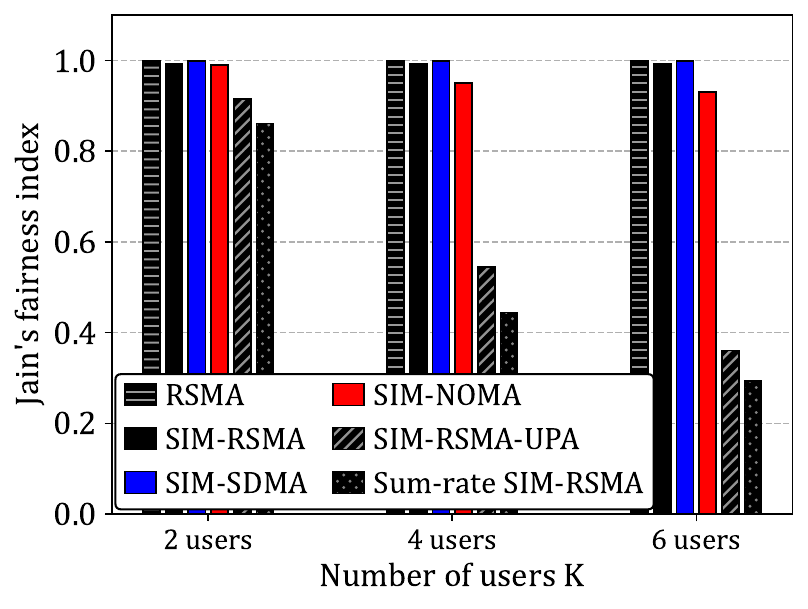}
        \vspace{-0.3in}
         \caption{The impact of the number of users $K$ on Jain's fairness index.}
        \label{fig:fairness_index}
    \end{minipage}
     \quad
     \begin{minipage}{0.28\textwidth}
         \centering
         \includegraphics[width=\textwidth]{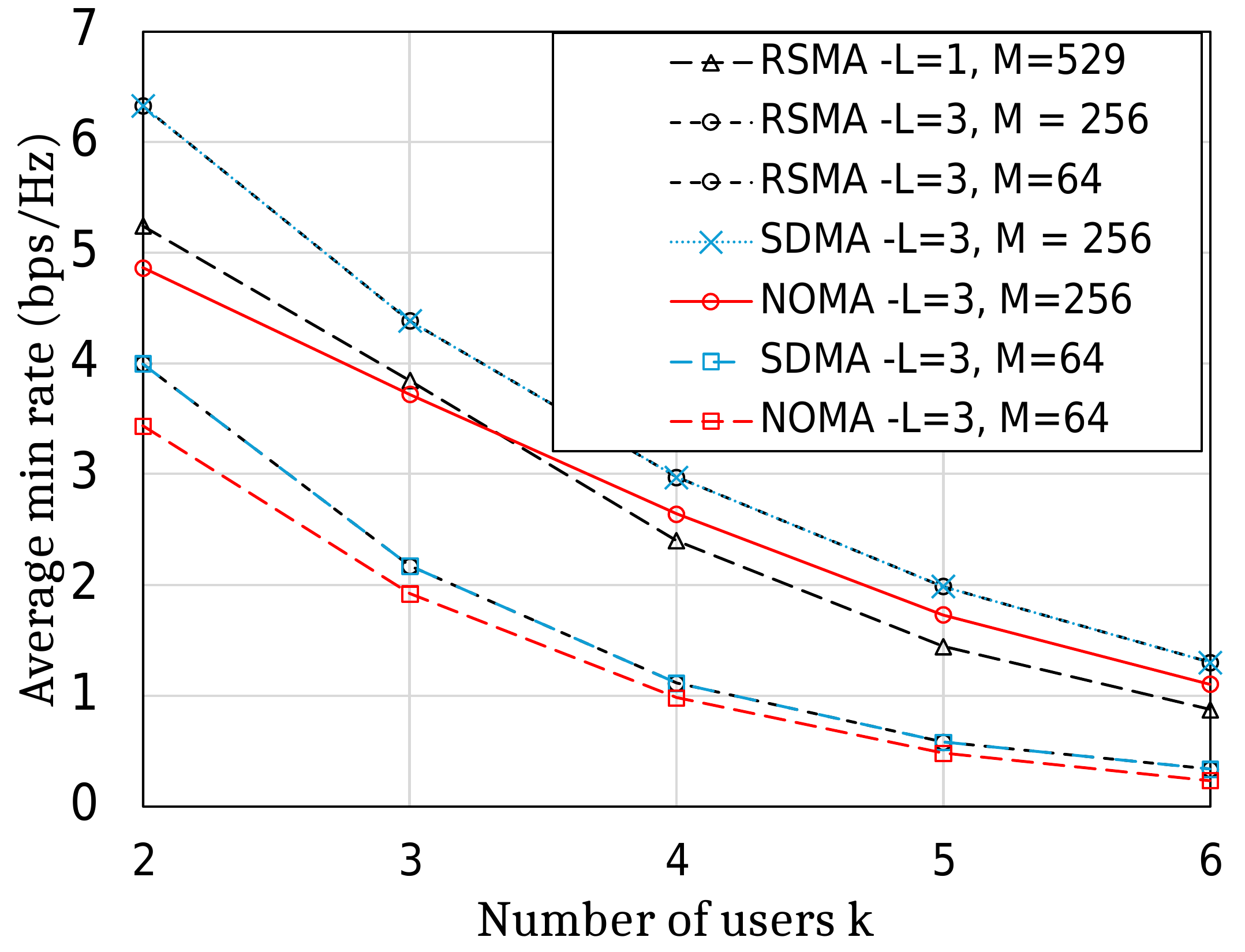}
        \vspace{-0.3in}
         \caption{The impact of the number of users $K$ on the minimum rate.}
         \label{fig:user}
     \end{minipage}
\end{figure*}

Fig.~\ref{fig:pmax} illustrates the impact of the transmit power $P_{\max}$ on the average minimum rate. The proposed SIM-RSMA scheme consistently outperforms SIM-NOMA across all power levels and achieves performance nearly identical to SIM-SDMA. At $P_{\max}=20\,\mathrm{dBm}$, SIM-RSMA achieves approximately {20\%} higher minimum rate than SIM-NOMA. The fully digital RSMA scheme provides an upper bound but at the cost of significantly higher hardware complexity and energy consumption. The similarity between SIM-RSMA and SIM-SDMA arises because, in the fully analog configuration, the BS is restricted to power allocation and lacks a digital precoder to shape the common-stream beam, while the SIM is constrained by unit-modulus phase shifts that prevent simultaneous alignment of the common signal toward all users. Since the common rate is limited by the weakest user, increasing it becomes disproportionately costly compared to enhancing private links, causing the optimization to naturally prioritize private streams and making SIM-RSMA effectively behave as a SIM-assisted SDMA configuration.

Figure~\ref{fig:fairness_index} evaluates the impact of varying the number of users $K$ on Jain's fairness index for different schemes. As shown, SIM-RSMA, SIM-SDMA, and the digital RSMA schemes maintain near-ideal fairness, with their indices around 0.99 as $K$ increases to 6 users, which demonstrates their robustness in ensuring equitable resource allocation. In contrast, the SIM-RSMA scheme with the sum-rate maximization strategy exhibits significantly poor fairness performance. This is due to the inherent conflict between maximizing aggregate throughput and ensuring equitable user rates. On the other hand, SIM-NOMA exhibits a decline in fairness as $K$ grows, which can be attributed to the constraints of power-domain multiplexing in maintaining multi-user fairness. Additionally, the performance of SIM-RSMA-UPA further emphasizes the necessity of optimized power allocation, as uniform power distribution results in significant fairness deterioration. These findings demonstrate that the SIM-RSMA scheme can achieve strong fairness performance without sacrificing scalability, while also providing additional rate improvements over SDMA and NOMA schemes. 

Fig.~\ref{fig:user} evaluates the impact of the number of users $K$ on the average minimum rate. As $K$ increases, all schemes experience a gradual decline due to stronger multiuser interference and reduced spatial degrees of freedom. SIM-RSMA and SIM-SDMA achieve similar performance for all $K$, while both outperform SIM-NOMA by approximately 20\% on average. With a small aperture ($M\!=\!64$), SIM-SDMA and SIM-NOMA exhibit nearly identical performance, reflecting limited spatial resolution and high channel correlation. Increasing the aperture to $M\!=\!256$ widens the performance gap, highlighting the importance of spatial resources. Furthermore, distributing $M\!=\!256$ meta-atoms over multiple layers yields around 40\% higher minimum rates than using a single-layer structure, as the multilayer configuration enables more effective spatial separation through successive wave transformations.

\begin{figure*}[t]
     \centering
     \begin{minipage}{0.3\textwidth}
         \centering
         \includegraphics[width=\textwidth]{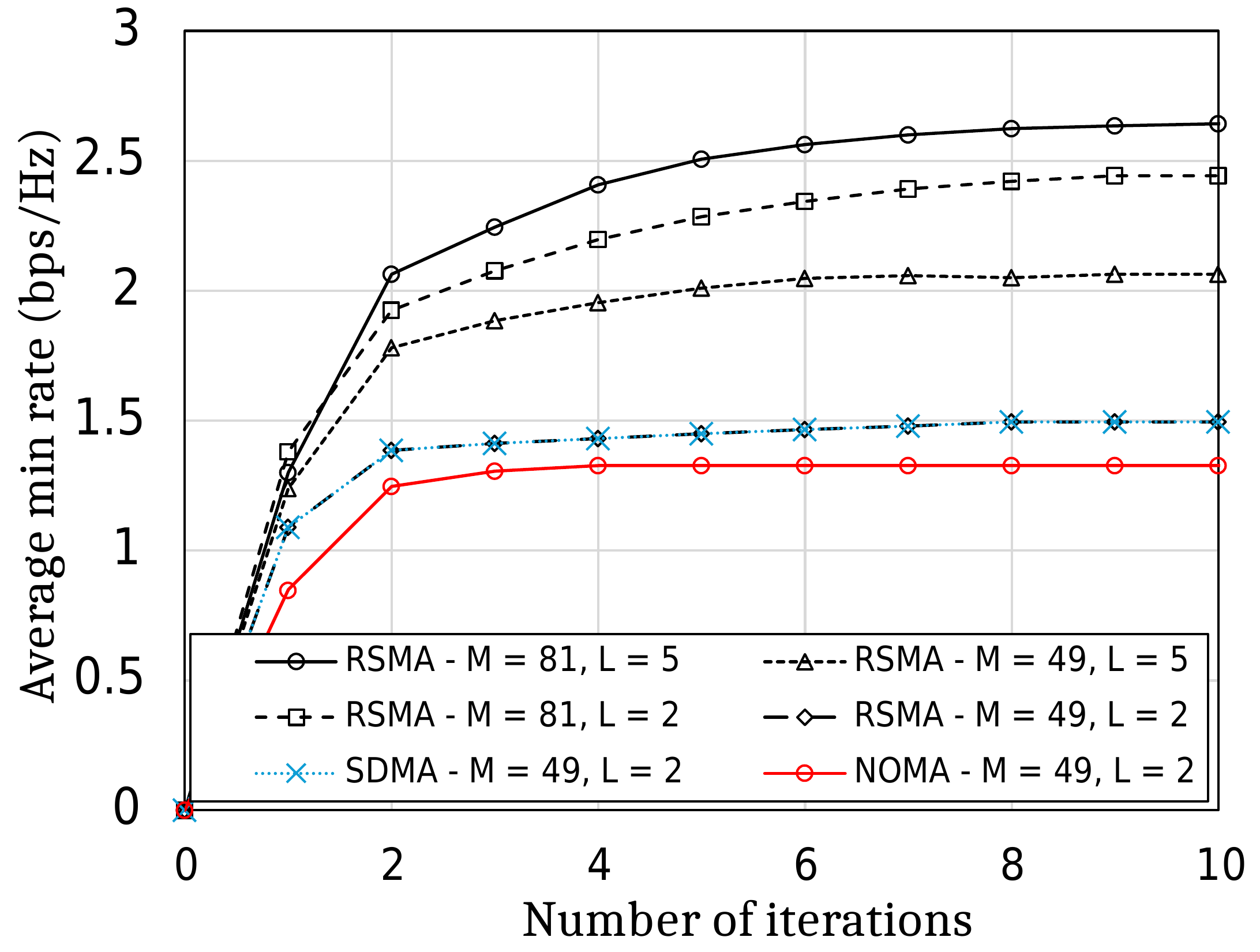}
        \vspace{-0.3in}
         \caption{Convergence behavior of the max-min AO algorithm in iterations.}
        \label{fig:iter}
    \end{minipage}
     \quad
     \begin{minipage}{0.3\textwidth}
         \centering
         \includegraphics[width=\textwidth]{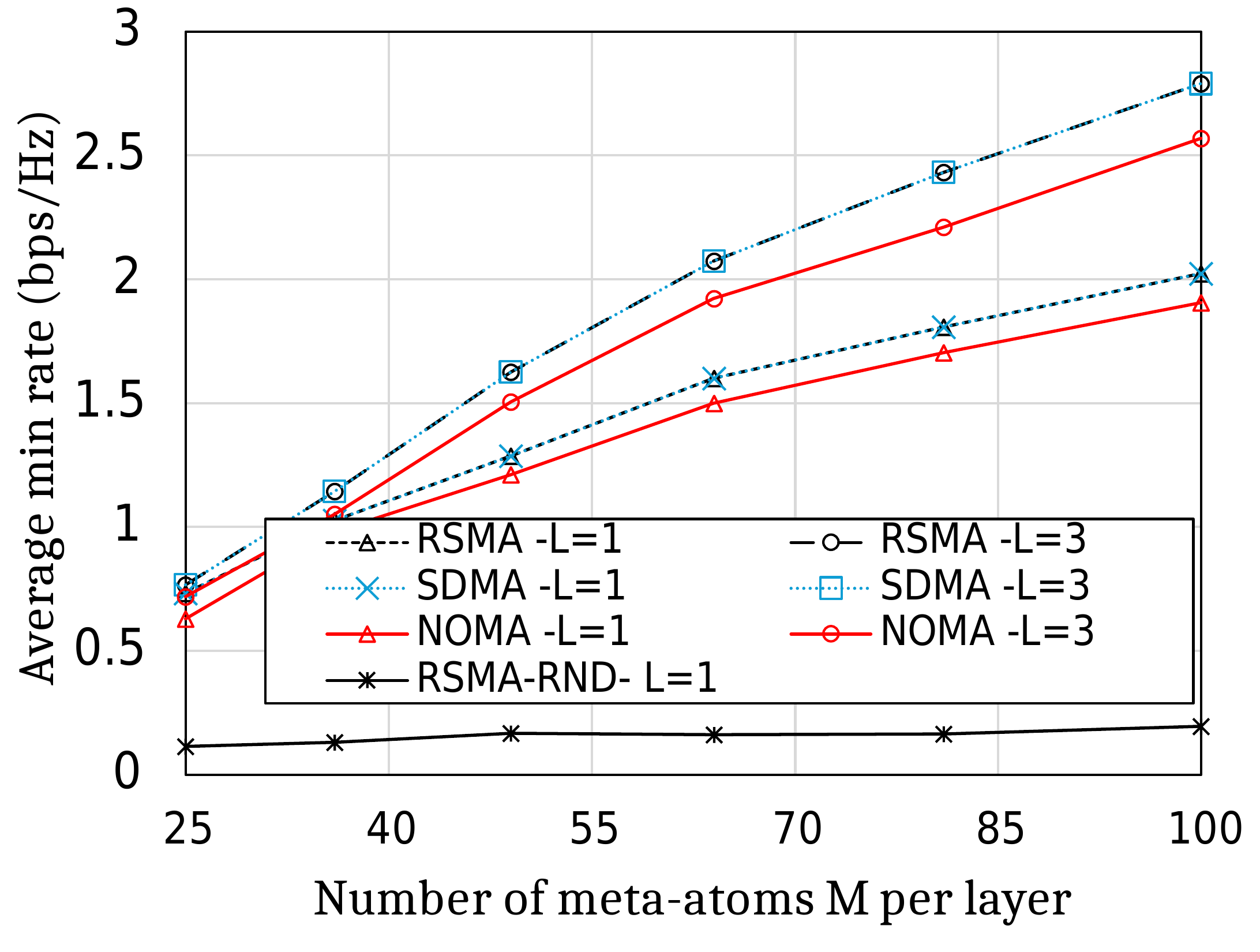}
        \vspace{-0.3in}
         \caption{The impact of the number of meta-atoms $M$ on the minimum rate.}
        \label{fig:m}
    \end{minipage}
     \quad
     \begin{minipage}{0.3\textwidth}
         \centering
         \includegraphics[width=\textwidth]{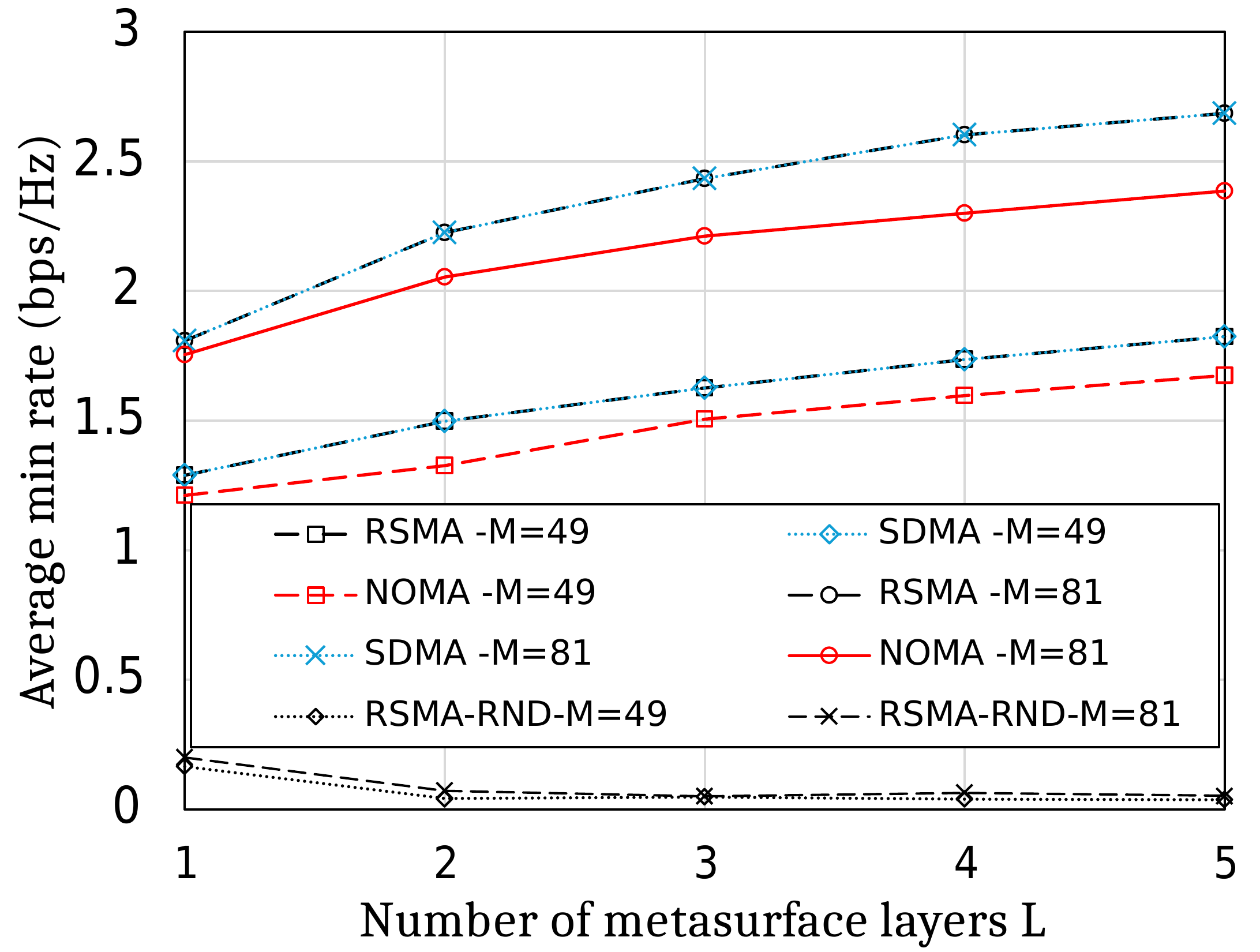}
        \vspace{-0.3in}
         \caption{The impact of the number of SIM layers $L$ on the minimum rate.}
         \label{fig:l}
     \end{minipage}
\vspace{-3mm}
\end{figure*}

Fig.~\ref{fig:iter} illustrates the convergence behavior of different schemes. All schemes converge rapidly within approximately 4--9 AO iterations, confirming the efficiency of the proposed optimization framework. For the same metasurface configuration, RSMA and SDMA achieve nearly identical convergence and minimum rates, while both outperform NOMA by about 15\% on average. Increasing $M\!=\!49$ to $M\!=\!81$ improves the minimum rate by 30\%, whereas increasing the $L\!=\!2$ to $L\!=\!5$ yields an additional 10\% gain. 

Figs.~\ref{fig:m} and~\ref{fig:l} examine the impact of the SIM's key architectural parameters on the average minimum rate: the number of meta-atoms per layer $M$, which defines the aperture size and spatial resolution, and the number of layers $L$, which determines the depth of wave transformations. RSMA and SDMA achieve nearly identical performance across all configurations, both consistently outperforming NOMA. Increasing $M$ provides a substantial improvement due to enhanced spatial focusing, whereas increasing $L$ yields smaller gains through deeper wave transformations. Specifically, increasing the aperture from $M\!=\!25$ to $M\!=\!100$ results in a {297\%} improvement in the minimum rate, while increasing the number of layers from $L\!=\!1$ to $L\!=\!5$ yields a {53\%} gain. These results indicate that once sufficient beamforming capability is achieved, expanding the aperture size $M$ is generally more effective than adding further layers $L$ in efficient interference suppression.

\vspace{-3mm}

\section{Conclusion}

In this paper, we investigated the integration of RSMA with a SIM for downlink multiuser MISO systems under a max–min fairness design. By leveraging SIM’s programmable wave-domain beamforming and the RSMA transmission strategy, we formulated and efficiently solved a max–min rate optimization problem using an alternating optimization framework, where power allocation was handled through successive convex approximation and SIM beamforming was optimized via the Riemannian conjugate gradient method. Simulation results showed that the proposed SIM-RSMA framework achieves performance nearly identical to SIM-SDMA while outperforming SIM-NOMA in terms of minimum user rate and fairness. These results indicate that, in the fully analog regime, the benefits of rate-splitting under max–min fairness are constrained by the shared SIM architecture and its unit-modulus phase-shift limitation, which restricts the independent shaping of the common stream and causes the system to effectively behave as SDMA. Therefore, future research should investigate hybrid or digitally assisted SIM-RSMA architectures that combine wave-domain beamforming with limited digital precoding, enabling user-specific control and fully exploiting RSMA’s interference management capability under fairness-driven objectives. 
\label{sec:conc}
\vspace{-1mm}

\bibliographystyle{IEEEtran}
\bibliography{bibliography.bib}

\end{document}